# General relativistic manifestations of orbital angular and intrinsic hyperbolic momentum in electromagnetic radiation.

J Strohaber

Department of Physics, Florida A&M University, Tallahassee, FL 32307, USA

E-mail: jstroha1@gmail.com

**Abstract**
General relativistic effects in the weak field approximation are calculated for electromagnetic Laguerre-Gaussian (LG) beams. The current work is an extension of previous work on the precession of a spinning neutral particle in the weak gravitational field of an optical vortex. In the current work, the metric perturbation is extended to all coordinate configurations and includes gravitational effects from circular polarization and intrinsic hyperbolic momentum. The final metric reveals frame-dragging effects due to intrinsic spin angular momentum (SAM), orbital angular momentum (OAM), and spin-orbit (SO) coupling. When investigating the acceleration of test particles in this metric, an unreported gravitational phenomenon was found. This effect is analogous to the motion of charged particles in the magnetic field produced by a current carrying wire. It was found that the gravitational influence of SAM and OAM affects test-rays traveling perpendicular to the intense beam and from this a gravitational Aharonov-Bohm analog is pursued.

## 1. Introduction

Gravity is a familiar force experienced by us daily: it is the force that holds us to the earth's surface, and it is the force that keeps the planets in orbit around our star. In 1687, Newton presented the first quantification of the gravitational force in his Principia and showed that the gravitational force between two massive bodies is proportional to the production of their masses and inversely proportional to the square of the distance between them [1]. While Newton's university law quantified the gravitational force, it did not provide a description of gravity, and it was not until 1915, when Einstein discovered general relativity (GR), that a more in-depth understanding of gravity as being due to the geometry of spacetime was given [2]. This description of gravity lead to new phenomena such as black holes, closed timelike curves, and gravitational waves and lensing. In Einstein's field equations, the geometry of spacetime is related to the energy-momentum configuration of a source and since light possesses energy and momentum, it too is expected to produce a gravitational field. In 1931, Tolman et. al. theoretically pursued such a task by calculating general relativistic effects from a beam of light beam [3].

General relativistic effects produced by beams of light have received renewed interest in recent years [4-9]. Interest in this field of research is motivated by a number reasons: curiosity, recent advancements in laser technology [10,11]; and experimental devices employed to detected weak general relativist effects such as gravitational waves [12,13], and geodetic and frame-dragging precessions [14]. Currently, for all measurable general relativistic phenomena, experimentalist have played the role of the observer and not the role of the originator of such gravitational perturbations [12—15]. The reason for this is clear when considering the magnitudes of gravitational perturbations far away from massive bodies such as the sun.

To get an idea of the strength of general relativistic effects in the vicinity of a powerful laser beam, the metric perturbation $h = \kappa \rho_L$ (where $\kappa$ is Einstein's constant, and $\rho_L$ is the linear energy density) embedded in a flat background Minkowski spacetime can be estimated [4—9] for modern high-powered lasers such as the National Ignition Facility (NIF) and Hercules systems [10,11]. The NIF laser system delivers 6 meter-long pulses each having an energy of 4MJ resulting in a linear energy density of $\rho_L \approx 6.7 \times 10^5 \, \text{J/m}$. Multiplying this energy density by Einstein's constant $\kappa \approx 2.1 \times 10^{-43} \, \text{m} \cdot \text{J}^{-1}$ gives a metric perturbation on the order of $h \approx 10^{-37}$. For comparison, the Advanced Laser Interferometer Gravitational-Wave Observatory (LIGO) has a detection sensitivity on the order of $h \approx 10^{-23}$ and implies that the NIF laser system would need to deliver $10^{14} \, \text{MJ}$ more energy per pulse to produce an "observable effect" [12]. The Hercules laser system produces a similar order of magnitude in the metric perturbation but on a timescale of 30 fs [11]. Another way to get a feel for the gravitational influence of a laser beam is by its mass-equivalence. The mass-equivalence of light pulses from the NIF laser is about $\sim 10^{-5} \, \text{mg}$ and that of the Hercules laser is about $\sim 10^{-10} \, \text{mg}$. As a comparison, the mass of a grain of sand is about 10 mg. One advantage of light, however, is that its "mass density" can be readily increased through focusing. Focusing pulses of light from the NIF and Hercules lasers to about a micron achieves a "mass density" of $\sim 0.6 \, \text{mg/cm}^3$ and about $\sim 6 \, \text{mg/cm}^3$ respectively. As a comparison, the least dense material produced by scientists is graphene aerogel which has a density of $\sim 0.2 \, \text{mg/cm}^3$ [16]. From this analysis, it can be concluded that current laser light is "light".

Despite these estimates, various experimental schemes have been theoretically investigated that favor longer interaction times and interference phenomena. In 1979, Scully considered an experimental scheme consisting of co-propagating laser pulses with a weak probe beam and an intense subluminal beam [17]. A general relativistic treatment of this thought experiment showed that the probe beam would be deflection by the intense beam while undergoing a phase shift. Estimates of an interaction length of $10^6 \, \text{km}$ showed that the probe beam experienced a deflection of $10^{-2} \lambda$ and a phase shift of $10^{-20} \, \text{m}^{-1}$. In a more recent publication, Mallett et al. suggested using a neutron interferometer in the weak gravitational field of a ring laser. In this proposed experimental setup, the authors calculated a total phase shift of $10^{-32} \, \text{rad}$, which is beyond the sensitive of current neutron interferometric techniques of $10^{-13} \, \text{rad}$ [6].

In this paper, an extension of previous work on frame dragging from optical vortices is made [8]. Optical vortices are beams of electromagnetic radiation having a characteristic helicoidal phase front surrounding a point of undetermined phase analogous to a spiral staircase with the phase singularity at the point of the newel [18—23]. These beams are solutions to the paraxial wave equation (PWE) and possess a quantized amount of orbital angular momentum per photon equal to $\text{OAM} = \hbar \ell$. Currently, much research in singular optics has revolved around the production of OAM containing beams and the transfer of this physical quantity in its interaction with matter [24]. In the perturbative regime of quantum optics, one of the first experimental realizations of the transfer of OAM in a nonlinear process was in the second harmonic generation of optical vortices embedded in femtosecond radiation [25]. In this experiment, the transfer of OAM to second harmonic radiation was observed to follows an addition rule $\hbar \ell_{2\text{ND}} = 2\hbar \ell$ analogous to that found for frequency conversion. A series of recent experiments demonstrated a more complex transfer of OAM in the generation of Raman sideband [26—28]. This process was found to follow a now well-established OAM-algebra for Stokes and anti-Stokes orders and was definitively verified through phase measurements in a simultaneous Young double slit experiment. More recently, the transfer of OAM in the highly nonlinear process of high harmonic generation (HHG) was verified through several experiments by various groups [29—31].

This paper is organized as follows. Section 2 provides a mathematical description of circularly polarized Laguerre-Gaussian (LG) beams and their energy-momentum tensor. Section. 3 gives a brief description of Einstein's field equations in the weak field approximation, and the mathematical machinery for calculating the metric perturbation for these beams of light. In Sec. 4, the metric perturbation for LG beams are calculated, and in Secs. 4 and 5 the acceleration of massive particles and the velocities of test-rays in this spacetime are investigated.

## 2. Optical vortices

Optical vortices belong to a family of solutions of the paraxial wave equation (PWE) knowns more generally as Laguerre-Gaussian beams [13]. The mathematical form representing LG beams is given by,

$$\psi_\rho^\ell = N_{\rho,\ell} \left(\frac{\sqrt{2}r}{w}\right)^{|\ell|} L_\rho^\ell\left(\frac{2r^2}{w^2}\right) \exp\left[-\left(\frac{1}{w^2} - i\frac{k}{2R}\right)r^2\right] e^{i\ell\theta} e^{i(\rho+|\ell|+1)\arctan(z/z_0)}, \quad (1)$$

Here $\rho$ is the radial mode number (or hyperbolic momentum charge), $\ell$ is the orbital angular momentum mode number, $N^2 = 2\rho!/(\pi w^2 \Gamma(\rho+\ell+1))$ is the normalization constant, $w_0$ is the waist of the beam, $w = w_0\sqrt{1+z^2/z_0^2}$ is the beam spot size, $z_0 = kw_0^2/2$ is the Rayleigh range (characteristic length in z-direction), $R = z + z_0^2/z$ is the wavefront radius of curvature, $L_\rho^\ell$ are the Laguerre polynomials, $\exp(i\ell\theta)$ is the phase factor giving rise to the helicoidal wavefront, and $\Psi_G = (\rho+|\ell|+1)\arctan(z/z_0)$ is the Gouy phase. The paraxial wave equation is a scalar equation, and therefore Eq. 1 is a scalar solution. The associated electric and magnetic fields can be found from the vector potential $\vec{A} = \alpha\psi_\rho^\ell \hat{i} + \beta\psi_\rho^\ell \hat{j}$ and Maxwell equation under the Lorentz gauge [13]. These field quantities are providing in Appendix A. Here $\alpha$ and $\beta$ are polarization parameters such that the spin helicity is given by $\sigma_z = i(\alpha^*\beta - \alpha\beta^*)$. From the electric and magnetic fields, the Poynting vector associated with the solutions of Eq. 1 has the well-known form [32]

$$\vec{S} = c\rho_L \left[\frac{r}{R}\hat{e}_r + \frac{1}{k}\left(\frac{\ell}{r} - \sigma_z\frac{1}{2}\frac{\partial}{\partial r}\right)\hat{e}_\theta + \hat{e}_z\right]|\psi_\rho^\ell|^2. \quad (2)$$

The first term on the right is the radial flux density due to diffraction of the beam. The second term is the angular component which depends on the angular and radial mode numbers and spin helicity parameter, and the third term is the commonly encountered longitudinal flux density associated with plane waves. Integral curves of the vector field in Eq. 2 demonstrate a spiraling of the Poynting vector about the optical axis and is linked to orbital angular momentum of these beams.

As shown in Appendix A, the energy-momentum tensor associated with circularly polarized LG beams can be written as a sum of three energy-momentum tensors terms,

$$T_{\mu\nu} = \rho_L \left\{ \tau^P_{\mu\nu} + \tau^{SO}_{\mu\nu} \left[ \ell - \sigma_z \left( |\ell| - \frac{2r^2}{w^2} - \frac{4r^2}{w^2} \frac{L^{\ell+1}_{\rho-1}}{L^\ell_\rho} \right) \right] \frac{1}{kr} + \tau^D_{\mu\nu} \frac{r}{R} \right\} |\psi^\ell_\rho|^2. \qquad (3)$$

The first term in Eq. 3 corresponds to a diffraction-free plane electromagnetic wave with and non-uniform intensity profile $|\psi^\ell_\rho|^2$. The second terms contain elements which depend on the OAM and spin content of the beam, and the third term is the corresponds to the radial expansion of the beam due to diffraction effects. The tensors $\tau_{\mu\nu}$ showed in each term in Eq. 3 are equal to

$$\tau^{plane}_{\mu\nu} = \begin{bmatrix} 1 & 0 & 0 & -1 \\ 0 & 0 & 0 & 0 \\ 0 & 0 & 0 & 0 \\ -1 & 0 & 0 & 1 \end{bmatrix}, \qquad (4.1)$$

$$\tau^{SO}_{\mu\nu} = \begin{bmatrix} 0 & \sin(\theta) & -\cos(\theta) & 0 \\ \sin(\theta) & 0 & 0 & -\sin(\theta) \\ -\cos(\theta) & 0 & 0 & \cos(\theta) \\ 0 & -\sin(\theta) & \cos(\theta) & 0 \end{bmatrix}, \qquad (4.2)$$

$$\tau^{diff}_{\mu\nu} = \begin{bmatrix} 0 & \cos(\theta) & \sin(\theta) & 0 \\ \cos(\theta) & 0 & 0 & -\cos(\theta) \\ \sin(\theta) & 0 & 0 & -\sin(\theta) \\ 0 & -\cos(\theta) & -\sin(\theta) & 0 \end{bmatrix}. \qquad (4.3)$$

Equation 4.1 has been previously investigated by other groups and corresponds to the energy-momentum of a plane electromagnetic wave. Equation 4.2 was used in the work of Ref [8] to investigate the gravitational influence of orbital angular momentum on spinning test particle.

### 3. Einstein's equation and metric perturbation

Einstein's field equations describe the geometry of a spacetime in the presence of an energy-momentum distribution [3—6, 8, 17, 33]. We will not derive the weak-field equations since they can be found in most elementary texts on the topic. In brief, the geometry of spacetime is determined by the metric tensor $g_{\mu\nu}$ which in flat spacetime is equal to the Minkowski metric $\eta_{\mu\nu} = diag(+,-,-,-)$. In the weak-field approximation, the metric is taken as a flat background spacetime $\eta_{\mu\nu}$ with a small perturbation $h_{\mu\nu}$ embedded in it $g_{\mu\nu} = \eta_{\mu\nu} + h_{\mu\nu}$. Using the Hilbert gauge and the metric in the weak-field approximation, Einstein's field equations are found to reduce to,

$$\partial_\lambda \partial^\lambda h_{\mu\nu} = -\kappa \left( T_{\mu\nu} - \frac{1}{2} \eta_{\mu\nu} T \right) \qquad (5)$$

In Eq. 5, $T_{\mu\nu}$ is the energy-momentum tensor given in Eq. 3, and $T = T^\mu_{\ \mu}$ is its trace found by contraction with the Minkowski metric $\eta_{\mu\nu}$. By inspection of Eqs. 4, it can easily be seen that $T = 0$ in Eq. 5.

The energy-momentum tensor investigated in this work is due to a steady beam of light, and for this reason the metric perturbation in Eq. 5 is time-independent. The solution of Eq. 5 can, therefore, be found from the integral,

$$h_{\mu\nu}(x) = -\kappa \int G(\vec{x}, \vec{x}') T_{\mu\nu}(r', \theta', z') r' dr' d\theta' dz' . \tag{6}$$

Here $\kappa = 8\pi G / c^4$, and $G(\vec{x}, \vec{x}')$ is the modified cylindrical Green's functions. A compact form of the Green's function in cylindrical coordinates was given previous by [34], and it explicitly separates the angular coordinate from the radial and longitudinal coordinates,

$$G(x, x') = \frac{1}{\pi} \sqrt{\frac{1}{r'r}} \sum_{m=0}^{\infty} \epsilon_m \cos\left[m(\theta - \theta')\right] Q_{m-1/2}(\chi) . \tag{7}$$

Here $\epsilon_m$ is Neuman's factor which has the values $\epsilon_0 = 1$ and $\epsilon_{m \geq 1} = 2$. The last factor in Eq. 7 is the associated Legendre function of the second kind, and it has as its argument $2r'r\chi = r^2 + r'^2 + (z - z')^2$. These functions are given by,

$$Q_{m-1/2}(\chi) = \frac{\sqrt{\pi} \Gamma(m + 1/2)}{(2\chi)^{m+1/2} \Gamma(m+1)} {}_2F_1\left(\frac{2m+3}{4}, \frac{2m+1}{4}; m+1; \frac{1}{\chi^2}\right). \tag{8}$$

In Eq. 8, ${}_2F_1$ is the confluent hypergeometric function which is equal to,

$${}_2F_1\left(\frac{2m+3}{4}, \frac{2m+1}{4}; m+1; \frac{1}{\chi^2}\right) = \frac{\Gamma(m+1)}{\Gamma(m+1/2)} \sum_{n=0}^{\infty} \frac{\Gamma(m + 2n + 1/2)}{n! \Gamma(m+1+n)} \left(\frac{1}{\chi}\right)^{2n}. \tag{9}$$

Combining Eq. 8 and Eq. 9, yields the form of the Legendre function that will be used in this work,

$$Q_{m-1/2}(\chi) = \frac{1}{2^m} \sqrt{\frac{\pi}{2}} \sum_{n=0}^{\infty} \frac{\Gamma(2n + m + 1/2)}{n! 2^{2n} \Gamma(m+1+n)} \left(\frac{1}{\chi}\right)^{2n+m+1/2}. \tag{10}$$

To facilitate the presentation of further expressions that use Eq. 10, we will simplify it to read as $Q_{m-1/2} = \sum_{n=0}^{\infty} C_n^{(m)} (1/\chi)^{2n+m+1/2}$ with two special cases $C_0^{(0)} = \pi / \sqrt{2}$ and $C_0^{(1)}(\chi) = \pi\sqrt{2}/8$.

## 4. Angular, radial and longitudinal integration

The task of the next two sections will be partly focus on solving Eq. 6 for the metric perturbation of a circularly polarized LG beam. The integrals in Eq. 6 can be facilitated by performing the angular integral separately. This is possible because the angular dependence in the Green's function appears as a multiplicative factor $\cos[m(\theta - \theta')]$, and the energy-momentum tensor $T_{\mu\nu}$ has its angular parts in $\tau_{\mu\nu}^P$ and $\tau_{\mu\nu}^{SO}$ as separate multiplicative factors. The relevant integrals for all values of $m$ are,

$$\int_0^{2\pi} \cos[m(\theta - \theta')] \tau_{\mu\nu}^P d\theta' = \tau_{\mu\nu}^P \begin{cases} 2\pi & m = 0 \\ 0 & other \end{cases}, \tag{11.1}$$

$$\int_0^{2\pi} \cos[m(\theta-\theta')]\tau_{\mu\nu}^{SO}(\theta')d\theta' = \tau_{\mu\nu}^{SO}(\theta)\begin{cases}\pi & m=1\\ 0 & other\end{cases}. \tag{11.2}$$

Radial integration in Eq. 6 will be evaluated by two methods. The first method is a modification version of the thin shell approximation used in Ref [8] to obtain an analytical result, and the second method is numerical integration used for comparison with the analytical result. For calculations using the thin-shell approximation, it is found that the Dirac delta function must be modified since it leads to infinities along the optical axis, and it accidentally removes an intrinsic spin term from the metric perturbation. Thin-shell integration is accomplished by replacing the intensity profile of the beam with a delta function such that all radial integration variables $r'$ will be set to the radius of the shell upon integration [8].

$$\left|\psi_\rho^\ell\right|^2 \approx \frac{\delta(r'-r_{\rho\ell})}{2\pi r_{\rho\ell}}, \tag{12}$$

In Eq. 12, $r_{\rho,\ell}$ is the radius of the cylindrical shell found by calculating the radial "center of mass" of the beam. It is the adoption of this radius for the cylindrical shell that removes the infinity along the optical axis and restores a missing spin term. Calculations of the radial "center of mass" leads to a quite complicated expression for the radius of the cylindrical shell,

$$r_{\rho\ell} = \frac{w}{\sqrt{2}}\frac{\rho!\Gamma(1/2)\Gamma(\ell+1)}{\Gamma(\ell+1/2)\Gamma(\rho+1/2)\,_3F_2\left(-\rho,\ell+1/2,1/2;-\rho+1/2,\ell+1;1\right)} \tag{13}$$

Upon setting the hyperbolic momentum charge in Eq. 13 to zero $\rho=0$, the shell radius simplifies to $r_{0,\ell} = w2^{-1/2}\Gamma(\ell+1)/\Gamma(\ell+1/2)$ which for $\ell>0$ is near the peak of the doughnut-shaped intensity profile. For a Gaussian beam $\ell=0$, and the shell radius is equal $r_{0,0}=w/\sqrt{2\pi}$ which corresponds to 73% of maximum. Notice that if we had used the well-known formula for the radial position $r_{peak}=w\sqrt{\ell/2}$ corresponding to the exact peak in the intensity profile of an $LG_{\rho=0}^\ell$ beam, then the radial position of the peak for $\ell=0$ would have been $r_{peak}=0$. This leads to the metric perturbation blowing up at the origin and as will be shown shortly, numerical integration demonstrates that the metric perturbation is, in fact, finite at the origin. Using the radial "center of mass" avoids this problem and allows some use of it when dealing with beams having non-zero hyperbolic momentum charge $\rho\neq 0$.

Integration over the z-direction will be implied using a top-hat function $\Pi_{2z_0}(z')$ extending from $z=-L$ to $z=L$ so that the extent of the beam along the longitudinal direction is $2L$. The Green's function in Eq. 6 depends on $z'$ through $\chi$, and the energy-momentum tensor through the beam parameters $w(z')$, and $R(z')$; however, in this work, the beams will be non-diffracting so that their beam parameters are those at $z=0$. This non-diffracting approximation sets the beam size to a constant $w(z=0)=w_0$ and sets the third term in Eq. 3 for the energy momentum-tensor to zero since $R(z=0)\to\infty$. The remaining $z$ integration variable resides in the Green's function, and integration over the z-coordinate in Eq. 10 yields,

$$\int\left(\frac{1}{\chi}\right)^{2n+m+1/2}dz' = (z'-z)\left(\frac{2r'r}{r^2+r'^2}\right)^{2n+m+1/2}\,_2F_1^{(n,m)}\left(\frac{1}{2},2n+m+\frac{1}{2};\frac{3}{2};-\frac{(z'-z)^2}{r^2+r'^2}\right). \tag{14}$$

The confluent hypergeometric function in Eq. 14 has two special cases that will be of use later,

$$_2F_1^{(0,1)} = \sqrt{\frac{r^2 + r'^2}{r^2 + r'^2 + (z-z')^2}}, \tag{15.1}$$

$$_2F_1^{(0,0)} = \frac{\sqrt{r^2 + r'^2}}{(z'-z)} \ln\left|\frac{1}{\sqrt{r^2 + r'^2}}\left(\sqrt{r^2 + r'^2 + (z'-z)^2} + z' - z\right)\right|. \tag{15.2}$$

In Secs. 2—4 we have developed the mathematical machinery needed to calculate the metric perturbations of LG beams including spin, orbital angular momenta, and intrinsic hyperbolic momentum. In the next two sections, calculations of the metric perturbation will be performed.

## 5. Planewave term

Linear polarized electromagnetic beams having a delta function intensity profile centered on the optical axis of the beam have been invested by several authors when calculating the metric perturbation [3,4,17]. However, when including both OAM and hyperbolic momentum terms, a modification like that of Eqs. (12) and (13) must be made. The metric perturbation calculated in the section is for the first term of the energy-momentum tensor of Eq. 3. Substituting the Green's function of Eq. 6, and the modified Dirac delta function of Eq. 12, into Eq. 3 yields,

$$h_{\mu\nu}^P = -\kappa \rho_L \tau_{\mu\nu}^P \frac{1}{\pi} \sum_{m=0}^{\infty} \epsilon_m \int \sqrt{\frac{1}{r'r}} |\psi_\rho^\ell|^2 Q_{m-1/2} \cos\left[m(\theta - \theta')\right] r' dr' d\theta' dz' \tag{16}$$

Unless were needed for clarity, functional arguments will be suppressed. Angular integration of Eq. 16 can be readily carried out using the results of Eq. 11.1 which sets $m=0$ in the process. With this integration, Eq. 16 reduces to,

$$h_{\mu\nu}^P = -2\kappa \rho_L \tau_{\mu\nu}^P \int \sqrt{\frac{r'}{r}} |\psi_\rho^\ell|^2 Q_{-1/2} dr' dz' . \tag{17}$$

Integration over the $z$-direction is achieved by using the associated Legendre function given in Eq. 10, and the integral in Eq. 14 which results in

$$h_{\mu\nu}^P = -2\kappa \rho_L \tau_{\mu\nu}^P (z'-z) \sum_{n=0}^{\infty} C_n^{(0)} \int \sqrt{\frac{r'}{r}} |\psi_\rho^\ell|^2 \left(\frac{2r'r}{r^2 + r'^2}\right)^{2n+1/2} {}_2F_1^{(n,0)} dr' . \tag{18}$$

For space considerations, the bounds of integration along the $z$-direction in Eq. 18 have not yet been evaluated. Analysis of the expansion coefficients shows that the dominant term is the leading order term. Taking the $n=0$ term in the expansion and evaluating the integral bounds yields,

$$h_{\mu\nu}^P = -\kappa \rho_L \tau_{\mu\nu}^P 2\pi \int |\psi_\rho^\ell|^2 \ln\left|\frac{\sqrt{r^2 + r'^2 + (L-z)^2} + (L-z)}{\sqrt{r^2 + r'^2 + (L+z)^2} - (L+z)}\right| r' dr' . \tag{19}$$

The integral in Eq. 19 is the first main result of this section, and it will be numerical integrated along the radial coordinate and compared to analytical results. An analytical solution of Eq. 18 can be found by Integrating over the thin cylindrical shell in Eq. 12,

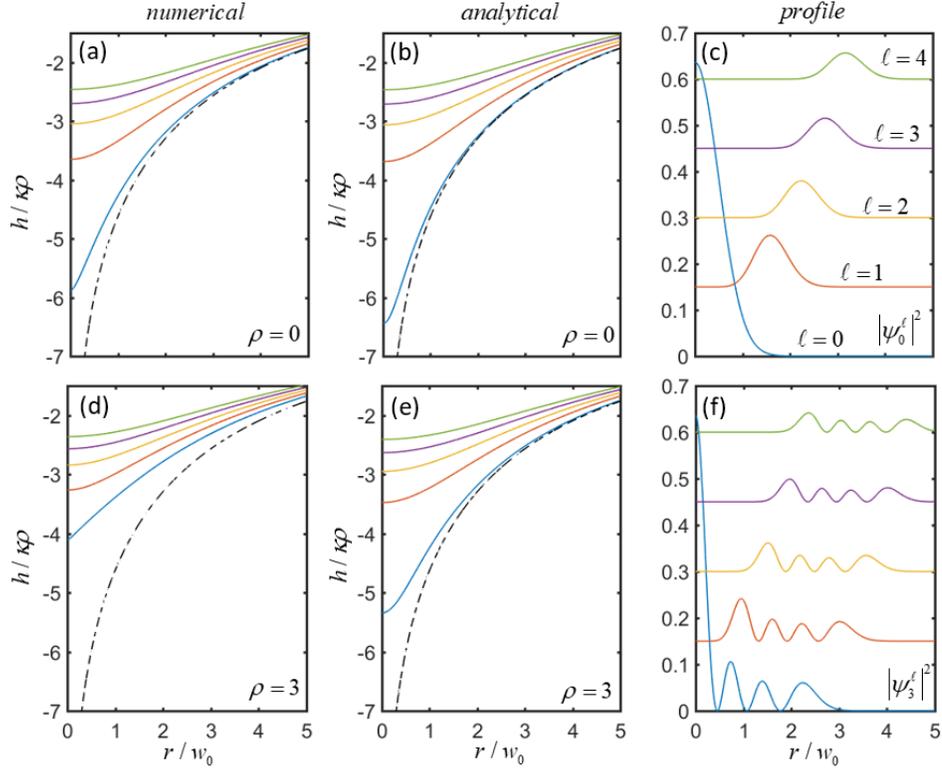

Fig. 1. (a) Numerically calculated scaled metric perturbation $h^P/\kappa\rho_L$ as a function of the radial distance $r/w_0$ from the optical axis and at $z=0$ for mode numbers $\ell=0,5,10,15,20$ and $\rho=0$. (b) Same as in (a) except the curves were calculated from the analytical solution of Eq. 21. (c) Modulus squared of the corresponding Laguerre Gaussian beams $|\psi_\rho^\ell|^2$ used for the curves in panels (a) and (b). (d)—(f) Same as (a)—(c) but with $\rho=3$. Dotting lines are the metric perturbation calculated for a thin pencil of light. This metric goes to negative infinity as $r$ goes to zero.

$$h^P_{\mu\nu} = -\kappa\rho_L\tau^P_{\mu\nu}\frac{(z'-z)}{\pi r_{\rho\ell}}\sum_{n=0}^{\infty}C_n^{(0)}\sqrt{\frac{r_{\rho\ell}}{r}}\left(\frac{2r_{\rho\ell}r}{r^2+r_{\rho\ell}^2}\right)^{2n+1/2} {}_2F_1^{(n,0)} \qquad (20)$$

To lowest order in the expansion of Eq. 20, and evaluating over the integral bound gives the second main result of this section,

$$h^P_{\mu\nu} = -\kappa\rho_L\tau^P_{\mu\nu}\ln\left|\frac{\sqrt{r^2+r_{\rho\ell}^2+(L-z)^2}+(L-z)}{\sqrt{r^2+r_{\rho\ell}^2+(L+z)^2}-(L+z)}\right| \qquad (21)$$

Equation 21 is comparable to solutions found by other authors with the exception that it is finite for all values of the radial coordinate $r$ due to the presence of $r_{\rho\ell}$ which is always greater than zero.

It will be convenient for plotting purposes to separate the tensorial part of the metric perturbation from the functional part as such $h^P_{\mu\nu} = \tau_{\mu\nu} h^P(r,z)$ with $h^P = h^P_{00}$. For comparison, the scaled-metric perturbation $h^P/\kappa\rho$ given in Eq. 19 and Eq. 21 are plotted in Fig. 1 as a function of the radial coordinate $r/w_0$ for angular mode numbers $\ell = 0, 5, 10, 15, 20$ and radial mode number $\rho = 0$ (top row) [$\rho = 3$ (bottom row)]. Analytical solutions of Eq. 21 are plotted in panels (b) and (e), and numerical solutions are plotted in panels (a) and (d). Panels (c) and (f) are plots of the modulus squared of the scalar amplitude function given in Eq. 1 as a function of the radial coordinate. In all calculations, the beam length parameter was taken to be $L = 5w$, and the longitudinal observation point was evaluated at $z = 0$. The dotted curves shown in columns 1 and 2 are calculated assuming a cylinder shell radius of $r_{\text{peak}} = w\sqrt{\ell/2}$ [8]. In this situation, the radius of the shell is zero when $\ell = 0$, and the beam reduces to an infinitesimally thin pencil of light. This thin pencil of light results in the metric perturbation diverging to infinity as one approached a transverse observation distance of $r = 0$. Numerical simulations demonstrate that the associated metric perturbation is, in fact, finite along the optical axis of a Gaussian beam and, therefore, using the radial "center of mass" in Eq. 13 removes this divergence. Analytical solutions show good agreement with numerical results for all mode numbers $\ell$, and $\rho$. The various curves can be identified by noting that the absolute values of the curve's amplitude decrease with the increase in $\ell$. For large observation points $r \gg r_{\rho\ell}$, the metric perturbation falls off as $h \sim -1/r$ as expected for the gravitational potential. This result is consistent with the gravitation potential around a massive cylindrical shell.

## 6. SAM and OAM terms

A metric perturbation endowed with orbital angular momentum near the optical axis of a beam has been previously investigated by Strohaber [8]. This metric perturbation was found to give rise to Lense-Thirring precession of a spinning neutral particle placed along the beam axis. The goal of this previous work was to investigate if OAM could result in a frame-dragging effect analogous to that found by Mallett for a ring laser. In this section, we extend upon the work of Strohaber by investigating the spacetime at all radial positions from a circularly polarized LG. The metric perturbation from the second term of the energy-momentum tensor including orbital angular momentum and spin can be written in short-hand notation as,

$$T^{SO}_{\mu\nu}(r,\theta) = \rho_L \tau^{SO}_{\mu\nu}(\theta') B^\ell_\rho(r';\sigma_z) \frac{1}{kr'} |\psi^\ell_\rho|^2, \tag{22}$$

where

$$B^\ell_\rho(r';\sigma_z) = \ell - \sigma_z |\ell| + \sigma_z \frac{2r'^2}{w^2} + \sigma_z \frac{4r'^2}{w^2} \frac{L^{\ell+1}_{\rho-1}}{L^\ell_\rho}. \tag{23}$$

Individual terms in Eq. 23 will be referenced as $B^\ell_\rho(r;\sigma_z) = B_\ell + B_{\sigma\ell} + B_\sigma + B_{\sigma\ell\rho}$ respectively. The first term on the right is the orbital angular momentum term, the second term is the spin-orbit term, the third term is the spin term, and the last term is the hyperbolic term [35, 36]. Substituting Eq. 7 for the Green's function, and Eq. 22 for energy-momentum tensor $T^{SO}_{\mu\nu}$ into Eq. 6 yields the following integral,

$$h^{SO}_{\mu\nu} = -\kappa\rho_L \frac{1}{k\pi} \sum_{m=0}^{\infty} \epsilon_m \int \sqrt{\frac{1}{r'r}} \cos\left[m(\theta-\theta')\right] \tau^{SO}_{\mu\nu} Q_{m-1/2} B^\ell_\rho |\psi^\ell_\rho|^2 \, dr' d\theta' dz'. \tag{24}$$

In contrast to Eq. 16, here both the Green's function and the energy-momentum tensor depend on the angular coordinate $\theta'$. The relevant integral is given in Eq. 11.2 and upon performing angular integration, the sum in Eq. 24 reduces to a single term with $m=1$,

$$h_{\mu\nu}^{SO} = -\kappa \rho_L \tau_{\mu\nu}^{SO} \frac{2}{k} \int \sqrt{\frac{1}{r'r}} Q_{1/2} B_\rho^\ell |\psi_\rho^\ell|^2 \, dr'dz'. \tag{25}$$

By making use of Eq. 10 for the associated Legendre function, Eq. 25 becomes

$$h_{\mu\nu}^{SO} = -\kappa \rho_L \tau_{\mu\nu}^{SO} \frac{2}{k} \sum_{n=0}^{\infty} C_n^{(1)} \int \sqrt{\frac{1}{r'r}} \left(\frac{1}{\chi}\right)^{2n+3/2} B_\rho^\ell |\psi_\rho^\ell|^2 \, dr'dz'. \tag{26}$$

Integrate over the extent of the beam along the $z$-direction and using the integral of Eq 14 yields,

$$h_{\mu\nu}^{SO} = -\kappa \rho_L \tau_{\mu\nu}^{SO} (z'-z) \frac{2}{k} \sum_{n=0}^{\infty} C_n^{(1)} \int \sqrt{\frac{1}{r'r}} \left(\frac{2r'r}{r^2+r'^2}\right)^{2n+3/2} {}_2F_1^{(n,1)} B_\rho^\ell |\psi_\rho^\ell|^2 \, dr'. \tag{27}$$

As before, the evaluation of the integral bounds has been postponed. Taking the lowest order term in the expansion $n=0$, and using Eq. 15.2 for the confluent hypergeometric function, Eq. 27 we find,

$$h_{\mu\nu}^{SO} = -\kappa \rho_L \tau_{\mu\nu}^{SO} \frac{\pi}{k} \int \frac{r'r}{r^2+r'^2} \left[ \frac{L-z}{\sqrt{r^2+r'^2+(z-L)^2}} + \frac{L+z}{\sqrt{r^2+r'^2+(z+L)^2}} \right] B_\rho^\ell |\psi_\rho^\ell|^2 \, dr'. \tag{28}$$

Equation. 28 is the first main result of this section and will be numerically integrated for comparison with analytical results. An analytical approximation to Eq. 27 can be obtained by integrating over the thin cylindrical shell in Eq. 12,

$$h_{\mu\nu}^{SO} = -\kappa \rho_L \tau_{\mu\nu}^{SO} \frac{(z'-z)}{k\pi r_{\rho\ell}} \sum_{n=0}^{\infty} C_n^{(1)} \sqrt{\frac{1}{r_{\rho\ell} r}} \left(\frac{2r_{\rho\ell} r}{r^2+r_{\rho\ell}^2}\right)^{2n+3/2} {}_2F_1^{(0,1)}(r_{\rho\ell}) B_\rho^\ell(r_{\rho\ell}). \tag{29}$$

To lowest order in the expansion and using Eq. 15.1 for ${}_2F_1^{(0,1)}$, Eq. 29 reduces takes the final form,

$$h_{\mu\nu}^{SO} = -\kappa \rho_L \tau_{\mu\nu}^{SO}(\theta) B_\rho^\ell(r_{\rho\ell}) \frac{r}{r^2+r_{\rho\ell}^2} \frac{1}{2k} \left[ \frac{L-z}{\sqrt{r^2+r_{\rho\ell}^2+(z-L)^2}} + \frac{L+z}{\sqrt{r^2+r_{\rho\ell}^2+(z+L)^2}} \right]. \tag{30}$$

Equation 30 is a more general version of the metric perturbation used in Ref [8]. By taking $\sigma_z = 0$ and for observation point close to the optical axis $r \ll r_0$, Eq. 30, gives the metric perturbation used in Ref [8] to investigate frame dragging from orbital angular momentum.

Analytical and numerical solutions of the metric perturbation $h_{\mu\nu}^{SO}$ are plotted in Fig. 2 for various values of $\sigma_z$, $\ell$ and $\rho$. In the First two rows $\sigma_z = -1$, $\rho=0$ and $\ell = 0,2,4,6,8,10$; and in the third and fourth rows $\sigma_z = -1$, $\rho = 3$ and $\ell = 0,2,4,6,8,10$. The columns are labeled as $B_\ell = \ell$, $B_{\sigma\ell} = -\sigma_z |\ell|$, $B_\sigma = \sigma_z 2r^2/w^2$, $B_{\sigma\ell\rho} = \sigma_z 4r^2 L_{\rho-1}^{\ell+1}/(w^2 L_\rho^\ell)$, and $B_\rho^\ell$ for the sum of these terms. In rows 1 and 3 are plotted the metric perturbations found from numerical integration of Eq. 28, and rows 2 and 4 are those from the analytical solutions of Eq. 30. In general, the amplitudes of the curves increase with increasing angular mode number $\ell$. For $\ell = 0$ the curves for $B_\ell = \ell$ and $B_{\sigma\ell} = -\sigma_z |\ell|$ are always zero. An important case occurs for $B_\sigma = \sigma_z 2r^2/w^2$. If we had chosen the delta function to have a radius $r_{peak} = w\sqrt{\ell/2}$ corresponding to the peak of the beam, then for $\ell = 0$ the curve for $B_\sigma = \sigma_z 2r_\ell^2/w^2$ would have been zero as well; however, numerical simulations show that this is not the case due to other portions of the beam have a non-zero contribution to the metric perturbation. In rows 1 and 2, the curves for $B_{\sigma\ell\rho}$ are zero since when $\rho = 0$ so is $L_{\rho-1}^{\ell+1} = 0$. For the curves plotted in rows 3 and 4, the radial mode number is nonzero $\rho = 3$. Here the analytical curves in row 4 for $B_\ell$, $B_{\sigma\ell}$ and $B_\sigma$ reasonably reproduce the numerical curves in row 3; however, the analytical curves for $B_{\sigma\ell\rho}$ poorly agree with numerical results.

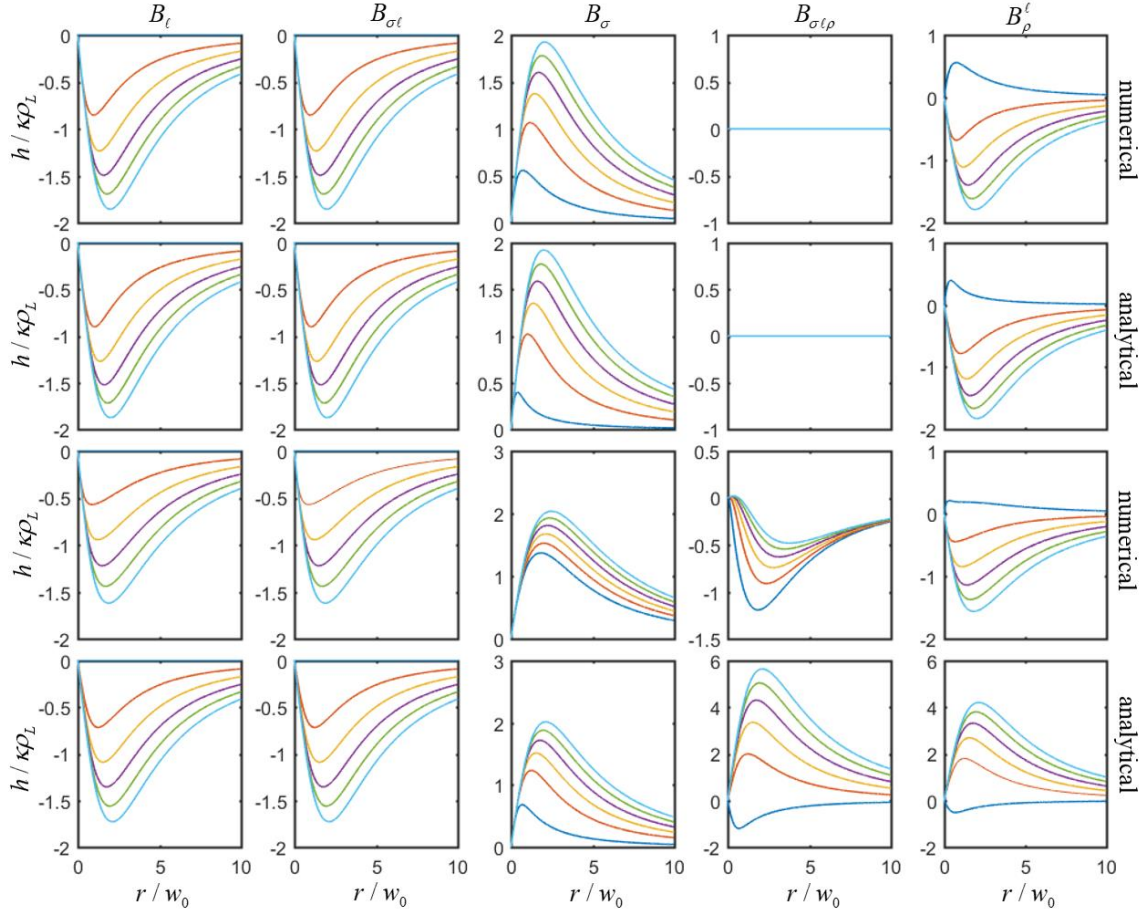

Fig. 2. Graphs of the metric perturbation $h^{SO}$ plotted for various values of $\ell$ and $\rho$. Data in the first and second rows are plotted for $\rho = 0$, $\sigma_z = -1$ and $\ell = 0, 2, 4, 6, 8, 10$; and data in the third and fourth rows are plotted for $\rho = 3$, $\sigma_z = -1$ and $\ell = 0, 2, 4, 6, 8, 10$. In general, the magnitude of the peaks of the curves increases with increasing $\ell$. In the first two column, the curves are zero $h^{SO} = 0$ when $\ell = 0$ due to $B_\ell = \ell$ and $B_{\sigma\ell} = -\sigma_z |\ell|$ terms. In column 3, the smallest amplitude curve is for $\ell = 0$. If $r_\ell = w\sqrt{\ell/2}$ had been used in the delta function approximation this curve would have been equal to zero, but numerical simulation shows it is not. In column 3, rows 3 and 4, the radial mode number is $\rho = 3$ and unlike the metric perturbation for the plane-wave part, the analytical solutions here show poor agreement with numerical calculations.

## 8. Particle Dynamics

Tolman has previously investigated the acceleration of stationary particles in the gravitational field of a pencil of light. Tolman's calculations showed that when a particle was placed midway between the ends of a pencil of light, the test particle experienced no acceleration along the beam axis. In this section, we calculate the general relativistic acceleration of a test particle beyond the stationary condition. The acceleration of a test particle in a spacetime can be found from the geodesic equation,

$$\frac{\partial^2 x^\mu}{\partial t^2} = -\Gamma^\mu_{\sigma\rho} \frac{\partial x^\sigma}{\partial t} \frac{\partial x^\rho}{\partial t} + \Gamma^0_{\sigma\rho} \frac{\partial x^\sigma}{\partial t} \frac{\partial x^\rho}{\partial t} \frac{\partial x^\mu}{\partial t} . \tag{31}$$

Equation 31 has been written in coordinate time and gives a set of coupled differential equations with terms linear, quadratic and cubic in the velocity. In the slow-motion approximation ($v \ll c$), terms quadratic and higher in velocity will be neglected. The interested reader is directed to Appendix C for a more in-depth analysis of all terms in Eq. 31. The nonzero terms on the right in Eq. 31 that are less than order $O(v^2)$ are,

$$\frac{\partial^2 x^\mu}{\partial t^2} = -\Gamma^\mu_{00} - 2\Gamma^\mu_{0i} \frac{\partial x^i}{\partial t} . \tag{32}$$

Here $2\Gamma^\mu_{\sigma\rho} = \eta^{\mu\nu}(h_{\sigma\nu,\rho} + h_{\rho\nu,\sigma} - h_{\sigma\rho,\nu})$ are the connection coefficients to lowest order in the perturbation $h_{\mu\nu}$ and reduce to $2\Gamma^\mu_{0\rho} = -(h_{0\mu,\rho} - h_{0\rho,\mu})$ due to the time-independence of $h_{\mu\nu}$. From this connection, the two properties $\Gamma^\mu_{0i} = -\Gamma^i_{0\mu}$ and $\Gamma^\mu_{0\mu} = 0$ can be readily seen. Equation 32 can be written more explicitly in matrix form as,

$$\begin{pmatrix} a_x \\ a_y \\ a_z \end{pmatrix} = -\begin{pmatrix} \Gamma^x_{00} \\ \Gamma^y_{00} \\ \Gamma^z_{00} \end{pmatrix} + \begin{pmatrix} 0 & -2\Gamma^x_{0y} & -2\Gamma^x_{0z} \\ 2\Gamma^y_{0y} & 0 & -2\Gamma^y_{0z} \\ 2\Gamma^x_{0z} & 2\Gamma^y_{0z} & 0 \end{pmatrix} \begin{pmatrix} v_x \\ v_y \\ v_z \end{pmatrix} . \tag{33}$$

Using the connection coefficients, Eq. 33 can be written in terms of the metric perturbation,

$$\begin{pmatrix} a_x \\ a_y \\ a_z \end{pmatrix} = -\frac{1}{2}\begin{pmatrix} h_{00,x} \\ h_{00,y} \\ h_{00,z} \end{pmatrix} + \begin{pmatrix} 0 & (h_{0x,y}-h_{0y,x}) & (h_{0x,z}-h_{0z,x}) \\ -(h_{0x,y}-h_{0y,x}) & 0 & (h_{0y,z}-h_{0z,y}) \\ -(h_{0x,z}-h_{0z,x}) & -(h_{0y,z}-h_{0z,y}) & 0 \end{pmatrix} \begin{pmatrix} v_x \\ v_y \\ v_z \end{pmatrix} . \tag{34}$$

From these equations, the acceleration is seen to have the same mathematical form as the Lorentz force found in electrodynamics. For this reason, the acceleration is written as $\vec{a} = \vec{\mathfrak{E}} + \vec{v} \times \vec{\mathfrak{B}}$, where $\vec{\mathfrak{E}} = (-\Gamma^x_{00}, -\Gamma^y_{00}, -\Gamma^z_{00})$ plays the role of the "electric" field, $\vec{\mathfrak{B}} = (-2\Gamma^y_{0z}, 2\Gamma^x_{0z}, -2\Gamma^x_{0y})$ that of the "magnetic" field, and $\vec{\mathfrak{A}} = (-h_{0x}, -h_{0y}, -h_{0z})$ and $\varphi = h_{00}/2$ play the role of the vector and scalar potentials respectively. With these correspondences, it can be easily shown that $\vec{\mathfrak{B}} = \vec{\nabla} \times \vec{\mathfrak{A}}$ and $\vec{\mathfrak{E}} = -\vec{\nabla}\varphi - \partial\vec{\mathfrak{A}}/\partial t$.

An analogy with the Lorentz force allows for easy interpretation of the influence of the metric perturbation on particle dynamics. In polar coordinates, the accelerations due to $\vec{\mathcal{E}}$ and $\vec{v} \times \vec{\mathcal{B}}$ in Eq. 33 and Eq. 34 are given by,

$$\vec{\mathcal{E}} = \frac{1}{2}\left(-\frac{\partial h^P}{\partial r}\hat{e}_r - \frac{\partial h^P}{\partial z}\hat{e}_z\right), \tag{35.1}$$

$$\vec{v} \times \vec{\mathcal{B}} = \frac{\partial h^P}{\partial r}v_z\hat{e}_r + \left(2\Gamma^x_{0y}v_r - \frac{\partial h^{SO}}{\partial z}v_z\right)\hat{e}_\theta - \frac{\partial h^P}{\partial r}v_r\hat{e}_z. \tag{35.2}$$

Equation 35.1 is the acceleration of test-particles towards the center of the intense beam due to its gravitational attraction. The physics of Eq. 35.2 is different from that of Eq. 35.1 and represents a velocity dependent acceleration. In Eq. 35.2, $2\Gamma^x_{0y} = -\partial h^{SO}/\partial r - h^{SO}/r$, and it can be seen from both equations that $h^P$ produces accelerations along radial $\hat{e}_r$ and longitudinal $\hat{e}_z$ directions, whereas in the angular direction only the metric $h^{SO}$ produces an acceleration.

For a particle moving only along the z-axis with radial coordinate $r=0$ and velocity $v=(0,0,z_z)$, the quantities $h^{SO}=0$, $\partial h^P_{00}/\partial r=0$ and $\partial h^{SO}/\partial z=0$ are zero and Eqs. 35 reduce to only the "electric" part,

$$\vec{\mathcal{E}} = -\frac{1}{2}\frac{\partial h^P}{\partial z}\hat{e}_z \tag{36}$$

Equation 36 shows that a particle placed along the optical-axis will experience and accelerated along the optical axis towards $z=0$. This can be seen from plots of the metric perturbations and its derivatives in Fig. 5 of Appendix B. When there is no SAM or OAM, that is when $\sigma=\ell=0$, and no restriction on the coordinate position of the test particle, then Eqs. 35 reduce to,

$$\vec{\mathcal{E}} = \frac{1}{2}\left(-\frac{\partial h^P}{\partial r}\hat{e}_r - \frac{\partial h^P}{\partial z}\hat{e}_z\right), \tag{37.1}$$

$$\vec{v} \times \vec{\mathcal{B}} = \frac{\partial h^P}{\partial r}(v_z\hat{e}_r - v_r\hat{e}_z). \tag{37.2}$$

In Eq. 37 the angular acceleration term is absent. In this case, the $\vec{\mathcal{E}}$ vector accelerates particles towards the center of the beam in both the radial and longitudinal directions. The vector $\vec{v}\times\vec{\mathcal{B}}$ produces rotational motion in any r-z plane. For example, assume an off-axis particle has an initial positive radial velocity $v_r$. Then by the second term of Eq. 37.2, the particle will experience an acceleration in the negative z-direction and thereby acquiring a component of velocity in the negative z-direction. The particle then (by the first term) experiences an acceleration in the negative radial direction. This results in an acceleration in the positive z-direction which results in a velocity in the positive z-direction producing an acceleration in the positive radial direction and so on. This dynamics is independent of SAM or OAM, and to the best of our knowledge has not been reported elsewhere.

Close to the optical axis $r \ll r_{\rho\ell}$ and in the plane $z=0$, Eqs. 35 reduces to,

$$\vec{v}\times\vec{\mathcal{B}} = 2\Gamma^x_{0y}v_r\hat{e}_\theta. \tag{38}$$

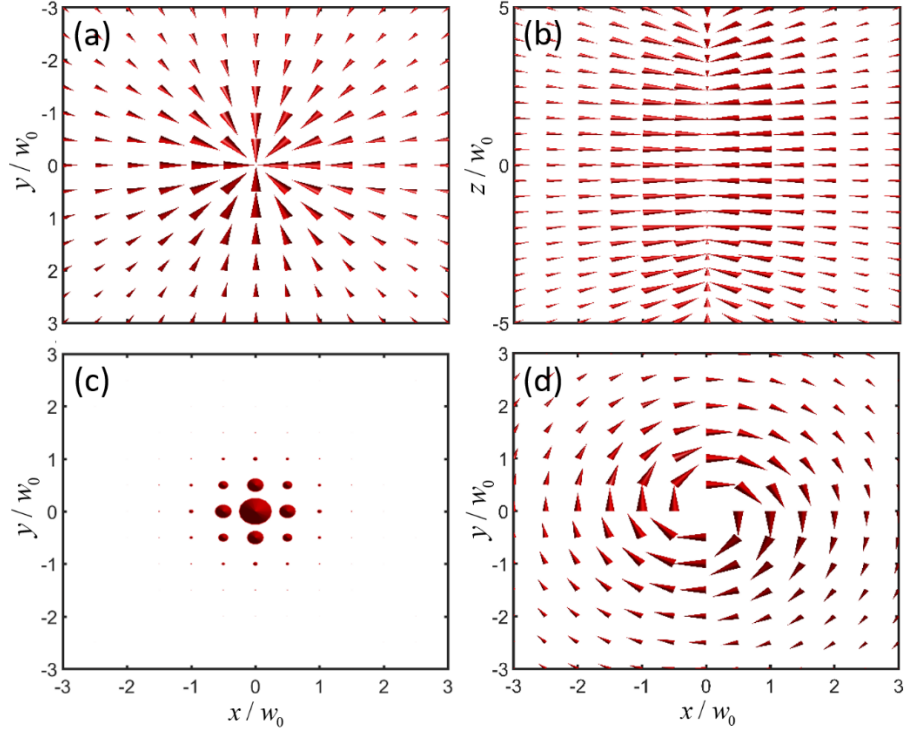

Fig. 3 "electric" and "magnetic" fields of the geodesic equation. (a) $x,y$-components of the "electric" field $\vec{\mathfrak{E}} = (-\Gamma_{00}^x, -\Gamma_{00}^y, 0)$ showing a radial gravitational acceleration towards the $r=0$. (b) $x,z$-components of the "electric" field $\vec{\mathfrak{E}} = (-\Gamma_{00}^x, 0, -\Gamma_{00}^z)$ showing gravitational acceleration towards the center of the beam $r=z=0$. (c) $x,y$-components of the "magnetic" field $\vec{\mathfrak{B}} = (-2\Gamma_{0z}^y, 2\Gamma_{0z}^x, 0)$ showing, by the right-hand rule, rotational acceleration $\vec{v} \times \vec{\mathfrak{B}}$ in radial planes where the acceleration is out of the page ($+z$) closer to the optical axis of the intense beam. (d) $z$-components of the "magnetic" field $\vec{\mathfrak{B}} = (0, 0, -2\Gamma_{0y}^x)$ out of the page and by the right-hand rule shows an acceleration in a counterclockwise direction. This acceleration direction changes with a change in sign of angular momentum mode number $\ell$.

Equation 38 shows that a particle having radial velocity will experience an angular deflection due to both SAM and OAM terms which appear in the factor $\Gamma_{0y}^x$. Thus far it has been shown that Eqs. 35 produce rotational motion in different planes and with different angular frequency. In fact, spectral analysis of the antisymmetric matrix in Eq. 34 shows that the eigenvalues are $\lambda_0 = 0$ and $\lambda_\pm = \pm i2\sqrt{(\Gamma_{0y}^x)^2 + (\Gamma_{0z}^x)^2 + (\Gamma_{0z}^y)^2}$. These eigenvalues are related to the angular frequencies of rotations in the various planes. For example, let the velocity dependent part of Eq. 34 be written as $\vec{a} = M \cdot \vec{v}$ and assuming a velocity vector $\vec{v} = \vec{v}_0 e^{i\omega t}$ and its acceleration $\vec{a} = i\omega \vec{v}$, the eigenvalue equation is $(M - i\omega)\vec{v} = 0$. From this, the rotational frequencies are $\omega_0 = 0$ and $\omega_\pm = \pm 2\sqrt{(\Gamma_{0y}^x)^2 + (\Gamma_{0z}^x)^2 + (\Gamma_{0z}^y)^2}$. Therefore, solutions to Eq. 38 are sinusoidal with an angular frequency of $\omega = 2\Gamma_{0y}^x$.

These rotational motions can be easily seen by viewing the test-particle in the metric perturbation as if it were a positively charged particle in the electromagnetic field surrounding a current carrying wire with an additional magnetic field in the direction of the current $\vec{\mathfrak{B}} = (0, 0, -2\Gamma^x_{0y})$. In Fig. 3 are plotted the components of these so-called "electric" and "magnetic" fields of the metric perturbation. In panel (a), the $x$ and $y$ components of the "electric" field $\vec{\mathfrak{E}} = (-\Gamma^x_{00}, -\Gamma^y_{00}, 0)$ indicate a radial acceleration towards the optical axis of the intense beam analogous to how a positively charged particle accelerates in the direction of electric fields. In panel (b), the $z$-component of the "electric" field $\vec{\mathfrak{E}} = (-\Gamma^x_{00}, 0, -\Gamma^z_{00})$ shows that test-particles are accelerated along the $z$-direction towards $z = 0$ as previously noted by Tolman. In panels (c) and (d), components of the "magnetic" field $\vec{\mathfrak{B}} = (-2\Gamma^y_{0z}, 2\Gamma^x_{0z}, -2\Gamma^x_{0y})$ can be used to infer the motion of test-particles by using the right-hand rule. For example, in Fig. 3(c) the "magnetic" field $\vec{\mathfrak{B}} = (0, 0, -2\Gamma^x_{0y})$ is out of the page in the positive $z$-direction. In this case, the fingers of the right-hand curl counterclockwise so the motion of a test particles will be clockwise. This motion is due to OAM since the plot is for $\sigma_z = 0$ and $\ell = -1$. In Fig. 3(d), the $x,y$-components of the "magnetic" field $\mathfrak{B}_x = -2\Gamma^y_{0z}$ and $\mathfrak{B}_y = 2\Gamma^x_{0z}$ are plotted, and the right-hand rule indicates rotational motion in radial planes. Rotations in the radial planes are independent of SAM and OAM, and when combined with the dragging effect from OAM and SAM results in a spiral motion around the optical axis.

## 9. TEST RAYS

Several authors have investigated the light-induced weak gravitational influence on test-rays [3, 17]. For example, Tolman found that test-rays traveling parallel and in the same direction of an intense beam experienced no change in their velocities while test-rays traveling in the opposite direction showed a variable speed. In this section, we are interested in the gravitational effects due to OAM and SAM on test-rays traveling in different directions relative to the intense beam. To allow for comparison with the work of Tolman, in this section, we have changed the signature of the Minkowski metric to $\eta_{\mu\nu} = (-, +, +, +)$ and multiplied our metric perturbations by negative one $-1$. In relativity, light travels along null trajectories given by the invariant distance $ds^2 = (\eta_{\mu\nu} + h_{\mu\nu})dx^\mu dx^\nu$. Using the metric perturbation and setting $ds$ to zero yields,

$$\left(1 - h^P_{00}\right) - v_x^2 - v_y^2 - \left(1 + h^p_{00}\right)v_z^2 = 2h^{SO}_{0x}v_x(1 - v_z) + 2h^{SO}_{0y}v_y(1 - v_z) - 2h^P_{00}v_z \tag{39}$$

For velocities of test-rays traveling parallel to the intense beam, the perpendicular velocities are set to zero $v_x = v_y = 0$ in Eq. 39, and after making use of the quadratic formula, Eq. 39 simplifies to,

$$v_z = \frac{h^p_{00} \pm 1}{1 + h^p_{00}}. \tag{40}$$

Equation 40 is the same result as obtained by Tolman. For test-rays traveling parallel to the beam in either direction, there are no relativistic effects of SAM or OAM from the intense beam. For test-rays traveling in the same direction ($+$), their velocity is that of the speed of light $v_z = +1$ regardless of the value of $h^p_{00}$. As pointed out by Tolman, this is a satisfying result for the stability of a beam of light [3]. For test-rays traveling in the opposite direction ($-$), their velocity is $v_z = -1$ when $h^p_{00} = 0$, and $v_z = 0$ when $h^p_{00} = 1$.

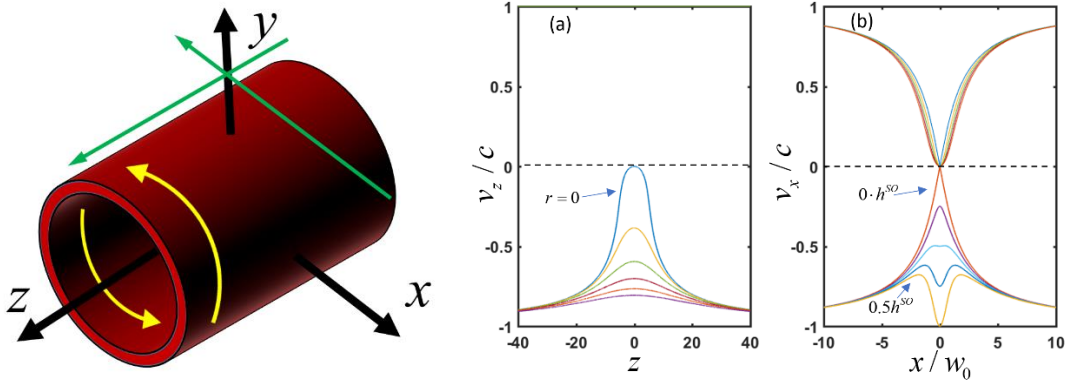

Fig. 4. For velocities in both panels (a) and (b), the maximum value of $h^P$ has been taken to be unity for demonstration purposes only. The beam parameters $\rho = 0$, $\sigma_z = 0$ and $\ell = 1$ were used in all calculations. In the upper illustration, the red cylinder represents the intense beam and the yellow arrows show the direction of the Poynting vector. (a) Plots of the longitudinal velocities of test-rays traveling parallel and in the same direction $v_z = (h_{00}^p + 1)/(1 + h_{00}^p)$ and in the opposite direction $v_z = (h_{00}^p - 1)/(1 + h_{00}^p)$ of the intense beam. In panel (a) when the test-ray is traveling in the same (positive) $z$-direction as the intense beam, its velocity is that of the speed of light $v_z = +1$, and test rays traveling in the opposition (negative) $z$-direction slow down within the region of the finite intense beam. In panel (a) these velocities are plotted for test-rays at various radial distances $r = [0,4,8,12,16,20]$ showing that the gravitational influence decreases with increases distance. In panel (b) velocities of test rays traveling perpendicular to the intense beams along the $x$-axis are plotted. Tests-rays raveling in the positive $x$-direction always slow down when passing by the intense beam; however, test-rays traveling in the negative $x$-direction can traveling at the speed of light when directly above the intense beam depending on the relative magnitudes of $h^{SO}$ and $h^P$.

In other words, when the metric perturbation is zero there is no gravitational effect and test-rays travel in the negative $z$-direction at the speed of light; however, when the metric perturbation is unity, the velocity of the speed of light is zero. As a reminder, the value of $h$ is extremely small $10^{-37}$ and the numbers given above have been used only to demonstrate the behavior of the test-rays.

The velocities of test-rays traveling perpendicular to the beam ($v_x$ and $v_y$) are found from Eq. 39 by setting $v_z = 0$ and $v_x = 0$ (or $v_y = 0$) and solving for the velocities in the $x$ and $y$ directions separately,

$$v_x = -h_{0x}^{SO} \pm \sqrt{\left(h_{0x}^{SO}\right)^2 + \left(1 - h_{00}^p\right)}$$
$$v_y = -h_{0y}^{SO} \pm \sqrt{\left(h_{0y}^{SO}\right)^2 + \left(1 - h_{00}^p\right)} \quad . \tag{41}$$

The results of Eq. 41 are different from those found by Tolman for the velocity of test-rays traveling in a direction perpendicular to the intense beam. The gravitational field due to intrinsic spin and external orbital angular momentum of the intense beam influences the speed of test-rays through $h^{SO}$. These velocities can have values between positive and negative $-1 \leq v \leq 1$ and by setting $v_x = 0$ and $v_y = 0$ separately, the following conditions $h_{0x}^{SO} = \pm h_{00}^p / 2$ and $h_{0y}^{SO} = \pm h_{00}^p / 2$ are found. As a reminder, we are in the weak-field approximation with the metric perturbation being much less than unity; however, for plotting purposes, we will set the maximum of $h_{00}^p$ to unity. In Fig. 4 is shown an illustration of the intense beams traveling in the positive $z$-direction, and the yellow arrows indicate the direction of the Poynting vector. The velocity curves of test-rays traveling parallel to the intense beam are shown in panel (a) as a function of the longitudinal direction. Each curve represents a test-ray traveling in the $z$-direction a distance $r = [0, 4, 8, 12, 16, 20]$ from the optical axis. For the curve with impact parameter of $r = 0$, the metric perturbation $h^P$ in Eq. 41 has been peak-normalized at $z = 0$ results in $v_z = 0$ at the origin. With increases impact parameter, the peak of the velocity curves decreases.

For test-rays traveling perpendicular to the intense beam, their velocity curves are plotted in panel (b) as a function of $x$. In the positive half of this plot is plotted $v_x = -h_{0x}^{SO} + \sqrt{(h_{0x}^{SO})^2 + (1 - h_{00}^p)}$ and in the negative half $v_x = -h_{0x}^{SO} - \sqrt{(h_{0x}^{SO})^2 + (1 - h_{00}^p)}$. For all curves $h^P$ in the positive half plane has been peak-normalized to unity. For velocities curves in the positive half plane, test-rays are traveling in the positive $x$-direction against the Poynting vector, and in the negative half of the plot, test rays are traveling in the negative $x$-direction with the Poynting vector. The different velocity curves are calculated by peak normalizing $h^P$ to unity with $h^{SO}$ peak normalized to [0.0, 0.125, 0.25, 0.375, 0.5]. When $h^{SO}$ is peak normalized to 0.5, the velocity of test-rays traveling with the azimuthal component of the Poynting (negative $x$-direction) as they pass directly above the intense beam is equal to the speed of light in the negative direction ($v_x = -1$). If the test-ray had been traveling in the negative $x$-direction below the intense beam (or if the sign of $\ell$ is changed) than the velocity curves in the upper in lower halves of the plot would be switched. This phenomenon is a analogous to a gravitational Aharonov-Bohm (GAB) effect where the metric perturbation plays the role of the electromagnetic potential. Although the measurability of this effect is expected to be beyond current detection limits, I will nevertheless briefly explore a design for a laboratory-scale experimental scheme.

An order of magnitude estimate can be made for a *Gedanken* experiment consisting of a long optical fiber wrapped into a cylindrical coil around the optical path of a high-powered laser beam in a cavity configuration. Two probe beams are sent in opposite direction around the coil and upon exiting allowed to interfere. Under the conditions $h_{0x}^{SO} = \pm h_{00}^p / 2$, one probe beam will travel at the speed of light in one direction around the coil while the other less than the speed of light in the other direction around the coil (Fig. 4) resulting in a phase difference upon exiting (I have neglecting dispersion effect in the fiber). The optical path difference (OPD) between two beams of light where one travels in a vacuum and the other in a medium with refractive index $n$ is given by $\Delta L / L \approx n - 1$ (where $n = c/v$). Substituting the condition $h_{0x}^{SO} = h_{00}^p / 2$ into Eq. 41 and expanding the result in a Taylor series yields

$$v_x = -h_{0x}^{SO} + \sqrt{(h_{0x}^{SO})^2 + (1 - 2h_{0x}^{SO})} \approx 1 - 2h_{0x}^{SO}. \tag{42}$$

Combining Eq. 42 with the OPD yields the simple result $\Delta L / L \approx h_{00}^p = 2h_{0x}^{SO}$. Here $h^P \approx \kappa \rho_L$ and so the OPD is equal to $\Delta L / L \approx 10^{-37}$. A strain of $\Delta L_{\text{LIGO}} / L_{\text{LIGO}} \approx 10^{-23}$ is measurable by Advanced LIGO and by setting $\Delta L = \Delta L_{\text{LIGO}}$ a fiber length of $L_f \approx 10^{14} L_{\text{LIGO}}$ ($L_{\text{LIGO}} = 4\text{km}$) is found for this thought experiment. The dimensions of the coil can be estimated by taking the volume of the fiber $V_f = \pi r_f^2 L_f$, and the volume of the coil $V_c = \pi r_c^2 L_c$ (taken as a cylinder) and setting them equal to each other $r_c^2 L_c = r_f^2 L_f$. Taking the fiber radius to be $r_f = 10\mu m$, and the coil length to be equal to ten times the radius of the coil $L_c = 10 r_c$, the coil radius and length are found to be equal to $r_c = 100$m and $L_c = 1000$m respectively. While these dimensions are terrestrial, there are numerous technical issues that render such an experiment unfeasible, and in all thought experiments of this type, the magnitude of $\kappa \rho_L$ effectively presents a fundamental limitation.

## 10. CONCLUSIONS

The metric perturbation of an intense Laguerre-Gaussian beam was calculated in the weak field approximation of general relativity. By using the radial "center of mass" of an LG beam, a modified Dirac delta function was used to avoid infinities at the origin that are typically encountered in these types of calculations. The results from this procedure agree well with results from numerical integration so that the extention of the metric perturbation to all radial coordinates are reliable when extracting physical information from the analytical results. Using these results, particle dynamics and velocities of test-rays were investigating. Both SAM and OAM were found to result in frame-dragging effects, and an unreported effect producing "gravitation eddies" in radial planes was found to accompany the beam independent of the SAM and OAM. Test-rays traveling parallel to the intense beam were found to be independent of SAM and OAM while test rays traveling perpendicular to the intense beam showed variable speeds that depended on SAM and OAM. This phenomenon was used to investigate a gravitational orbital angular momentum Aharonov-Bohm effect (GOAM-AB).

**Appendix A. Energy-Momentum Tensor**

The energy-momentum for a beam of electromagnetic radiation can be calculated from

$$T_{\mu\nu} = \begin{bmatrix} \frac{1}{2}(E^2 + B^2) & -S_x & -S_y & -S_z \\ -S_x & \sigma_{xx} & \sigma_{xy} & \sigma_{xz} \\ -S_y & \sigma_{yx} & \sigma_{yy} & \sigma_{yz} \\ -S_z & \sigma_{zx} & \sigma_{zy} & \sigma_{zz} \end{bmatrix}, \qquad (A.1)$$

where the Poynting vector and the Maxwell stress tensor are given respectively by,

$$\vec{S} = \frac{1}{\mu_0} \vec{E} \times \vec{B}, \qquad (A.2)$$

$$\sigma_{ij} = \varepsilon_0 E_i E_j + \frac{1}{\mu_0} B_i B_j - \frac{1}{2}\left(\varepsilon_0 E^2 + \frac{1}{\mu_0} B^2\right)\delta_{ij}. \qquad (A.3)$$

The electric and magnetic fields of a beam within the paraxial approximation can be found from Maxwell's equation as outlined in Ref [13] and are given by

$$\vec{E} = E_0\left[\alpha \hat{e}_x + \beta \hat{e}_y + \frac{i}{k}\left(\alpha \frac{\partial}{\partial x} + \beta \frac{\partial}{\partial y}\right)\hat{e}_z\right]|\psi|^2, \qquad (A.4)$$

$$\vec{B} = B_0 \left[ -\beta \hat{e}_x + \alpha \hat{e}_y - \frac{i}{k} \left( \beta \frac{\partial}{\partial x} - \alpha \frac{\partial}{\partial y} \right) \hat{e}_z \right] |\psi|^2. \tag{A.5}$$

From Eqs. A.4 and A.5, it can be seen that $B_y = E_x$ and $B_x = -E_y$. In calculating the energy-momentum tensor in the paraxial approximation using Eqs. A.2 and A.3, terms of second-order in wavelength are neglected (viz., $E_z^2, B_z^2 \approx 0$). These relations can be substituted into Eq. A.1 which reduces to,

$$T_{\mu\nu} = \begin{bmatrix} S_z & -S_x & -S_y & -S_z \\ -S_x & 0 & 0 & S_x \\ -S_y & 0 & 0 & S_y \\ -S_z & S_x & S_y & S_z \end{bmatrix}, \tag{A.6}$$

where in Cartesian coordinates the Poynting vector is given by

$$\vec{S} = |E_0|^2 \frac{1}{c\mu_0} \left[ \begin{array}{l} \left( \cos(\theta) \frac{r}{R(z)} - \sin(\theta) \frac{1}{k} \left( \frac{\ell}{r} - \sigma_z \frac{1}{2} \frac{\partial}{\partial r} \right) \right) \hat{e}_x \\ + \left( \sin(\theta) \frac{r}{R(z)} + \cos(\theta) \frac{1}{k} \left( \frac{\ell}{r} - \sigma_z \frac{1}{2} \frac{\partial}{\partial r} \right) \right) \hat{e}_y + \hat{e}_z \end{array} \right] |\psi|^2. \tag{A.7}$$

**Appendix B. Derivatives of the Metric Perturbation**

In this appendix, relevant derivatives of the metric perturbations $h_{\mu\nu}^P$ and $h_{\mu\nu}^{SO}$ are provided and plotted as a function of radial $r/w_0$ and longitudinal $z/w_0$ directions. To begin with, a common factor that appears frequently can be written in short-hand notation as

$$\begin{aligned} f_-(r,z) &= \sqrt{r^2 + r_{\rho\ell}^2 + g_-^2} \\ f_+(r,z) &= \sqrt{r^2 + r_{\rho\ell}^2 + g_+^2} \end{aligned}, \tag{B.1}$$

where $g_\pm = L \pm z$. When $z = 0$ these equations reduce to $g_- = g_+$ and $f_- = f_+$. With this handy notation, Eq. 21 and Eq. 30 can be written as follows

$$h_{\mu\nu}^P / \kappa \rho_L = -\tau_{\mu\nu}^P \ln \left| \frac{f_- + g_-}{f_+ - g_+} \right|, \tag{B.2}$$

$$h_{\mu\nu}^{SO} / \kappa \rho_L = -\tau_{\mu\nu}^{SO}(\theta) B_\rho^\ell (r_{\rho\ell}) \frac{1}{2k} \frac{r}{r^2 + r_{\rho\ell}^2} \left( \frac{g_-}{f_-} + \frac{g_+}{f_+} \right). \tag{B.3}$$

The radial and longitudinal derivatives of Eq. B.2 are

$$\frac{1}{\kappa \rho_L} \frac{\partial h_{\mu\nu}^P}{\partial r} = -\tau_{\mu\nu}^P r \left( \frac{1}{f_-(f_- + g_-)} - \frac{1}{f_+(f_+ - g_+)} \right), \tag{B.4}$$

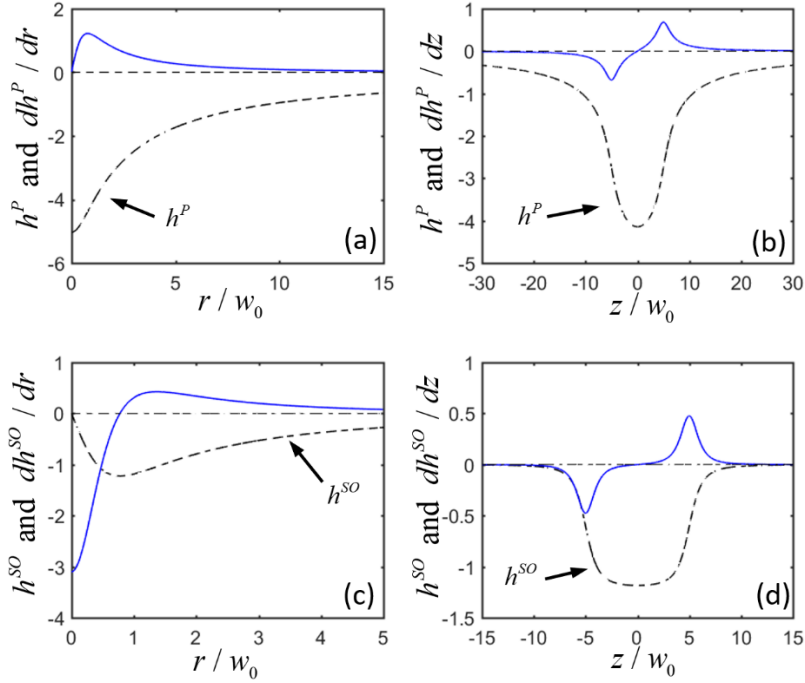

Fig. 5. (a) and (b) are curves of $h^P$ and their respective derivatives plotted as a function of (a) radial position $r$ and (b) longitudinal position $z$. (c) and (d) are the same as those for (a) and (b) except they are the curves for the metric perturbation $h^{SO}$ and its derivatives. The dotted curves are for $h^P$ and $h^{SO}$, and the solid blue curves are for their derivatives $dh^P/dr$, $dh^P/dz$, $dh^{SO}/dr$ and $dh^{SO}/dz$. In all plots, $w_0 = 1$, $z = 5w_0$, $\ell = 1$ and $\rho = 0$. Curves in panels (a) and (c) have a constant longitudinal position $z = 0$ and those in (b) and (d) have a constant radial position $r = w_0$.

$$\frac{1}{\kappa \rho_L} \frac{\partial h^P_{\mu\nu}}{\partial z} = -\tau^P_{\mu\nu}\left(\frac{1}{f_+} - \frac{1}{f_-}\right), \tag{B.5}$$

and those for Eq. B.3 are

$$\frac{1}{\kappa \rho_L} \frac{\partial h^{SO}_{\mu\nu}}{\partial z} = -\tau^{SO}_{\mu\nu}(\theta) B^\ell_\rho(r_{\rho\ell}) \frac{1}{2k} \frac{r}{r^2 + r_{\rho\ell}^2}\left(\frac{1}{f_+} - \frac{1}{f_-} - \frac{g_+^2}{f_+^3} + \frac{g_-^2}{f_-^3}\right), \tag{B.6}$$

$$\frac{1}{\kappa \rho_L} \frac{\partial h^{SO}_{\mu\nu}}{\partial r} = -\tau^{SO}_{\mu\nu}(\theta) B^\ell_\rho(r_{\rho\ell}) \frac{1}{2k} \frac{1}{r^2 + r_{\rho\ell}^2}\left[\left(1 - \frac{2r^2}{r^2 + r_{\rho\ell}^2}\right)\left(\frac{g_-}{f_-} + \frac{g_+}{f_+}\right) - r^2\left(\frac{g_-}{f_-^3} + \frac{g_+}{f_+^3}\right)\right]. \tag{B.7}$$

A few special cases should be noted. In the plane $z = 0$ the following quantities are zero: $\partial h^P/\partial r = \partial h^P/\partial z = 0$ and $\partial h^{SO}/\partial z = 0$, and when $r = 0$ we have $h^{SO} = 0$, $\partial h^P/\partial r = 0$ and $\partial h^{SO}/\partial z = 0$.

## Appendix C. Analysis of the Geodesic Equation

In this appendix, a derivation of velocity squared terms and higher of the geodesic of Eq. 31 equation are given. The geodesic equation in coordinate time is given by,

$$\frac{\partial^2 x^\mu}{\partial t^2} = -\Gamma^\mu_{00} - 2\Gamma^\mu_{0i}\frac{\partial x^i}{\partial t} - \Gamma^\mu_{ij}\frac{\partial x^i}{\partial t}\frac{\partial x^j}{\partial t} + \left(\Gamma^0_{00} + 2\Gamma^0_{0i}\frac{\partial x^i}{\partial t} + \Gamma^0_{ij}\frac{\partial x^i}{\partial t}\frac{\partial x^j}{\partial t}\right)\frac{\partial x^\mu}{\partial t}, \quad (C.1)$$

where the connection coefficients have been separated into time and spatial components. Using the connection coefficients $2\Gamma^\mu_{\sigma\rho} = \eta^{\mu\nu}(h_{\sigma\nu,\rho} + h_{\rho\nu,\sigma} - h_{\sigma\rho,\nu})$, the first term on the right can be written as,

$$\Gamma^\mu_{00} = \eta^{\mu\nu}\left(\frac{\partial h_{0\nu}}{\partial t} - \frac{1}{2}\frac{\partial h_{00}}{\partial x^\nu}\right) \to \frac{\partial \vec{\mathfrak{A}}}{\partial t} + \vec{\nabla}\varphi, \quad (C.2)$$

Where $\varphi = h_{00}/2$ and $\vec{\mathfrak{A}} = (-h_{0x}, -h_{0y}, -h_{0z})$. The second term in Eq C.1 is

$$2\Gamma^\mu_{0i}\frac{\partial x^i}{\partial t} = \eta^{\mu\nu}\left[\frac{\partial h_{i\nu}}{\partial t} + (h_{0\nu,i} - h_{0i,\nu})\right]\frac{\partial x^i}{\partial t} \to -\vec{v}\cdot\frac{\partial \vec{h}}{\partial t} - \vec{v}\times(\vec{\nabla}\times\vec{\mathfrak{A}}). \quad (C.3)$$

Here $\vec{\mathfrak{B}} = \vec{\nabla}\times\vec{\mathfrak{A}}$ and $\vec{h}$ a matrix equal to the spatial part of $h_{\mu\nu}$. Exception for the term $\vec{v}\cdot\partial\vec{h}/\partial t$, Eq. C.2 and Eq. C.3 are mathematically equivalent to the Lorentz force. The third term in Eq C.1 is somewhat more complex,

$$\Gamma^\mu_{ji}\frac{\partial x^j}{\partial t}\frac{\partial x^i}{\partial t} = -\frac{1}{2}(h_{i\mu,j} + h_{j\mu,i})\frac{\partial x^j}{\partial t}\frac{\partial x^i}{\partial t} + \frac{1}{2}h_{ij,\mu}\frac{\partial x^j}{\partial t}\frac{\partial x^i}{\partial t}$$
$$\Gamma^\mu_{ab}\frac{\partial x^a}{\partial t}\frac{\partial x^b}{\partial t} \to -\frac{1}{2}\left[2(\vec{v}\cdot\vec{\nabla}_h)(\vec{h}\cdot\vec{v}) - \vec{\nabla}_h(\vec{v}\cdot\vec{h}\cdot\vec{v})\right]. \quad (C.4)$$

The next two terms together are,

$$\Gamma^0_{00} + 2\Gamma^0_{0i}\frac{\partial x^i}{\partial t} = \frac{1}{2}\frac{\partial h_{00}}{\partial t} + \frac{\partial x^i}{\partial t}\frac{\partial h_{00}}{\partial x^i} = \frac{\partial \varphi}{\partial t} + 2(\vec{v}\cdot\vec{\nabla})\varphi, \quad (C.5)$$

and the last term is

$$\Gamma^0_{ij}\frac{\partial x^i}{\partial t}\frac{\partial x^j}{\partial t} = \frac{\partial x^i}{\partial t}\frac{\partial x^j}{\partial t}\frac{\partial h_{i0}}{\partial x^j} - \frac{1}{2}\frac{\partial x^i}{\partial t}\frac{\partial x^j}{\partial t}\frac{\partial h_{ij}}{\partial t}$$
$$\Gamma^0_{ij}\frac{\partial x^i}{\partial t}\frac{\partial x^j}{\partial t} = -(\vec{v}\cdot\vec{\nabla}_{\mathfrak{A}})(\vec{v}\cdot\vec{\mathfrak{A}}) - \frac{1}{2}\vec{v}\cdot\left(\frac{\partial \vec{h}}{\partial t}\cdot\vec{v}\right) \quad (C.6)$$